# Forecasting the onset and course of mental illness with Twitter data

Andrew G. Reece[a*], Andrew J. Reagan[b,c], Katharina L.M. Lix[d],
Peter Sheridan Dodds[b,c], Christopher M. Danforth[b,c*], Ellen J. Langer[a]

**Abstract:** We developed computational models to predict the emergence of depression and Post-Traumatic Stress Disorder in Twitter users. Twitter data and details of depression history were collected from 204 individuals (105 depressed, 99 healthy). We extracted predictive features measuring affect, linguistic style, and context from participant tweets (N=279,951) and built models using these features with supervised learning algorithms. Resulting models successfully discriminated between depressed and healthy content, and compared favorably to general practitioners' average success rates in diagnosing depression. Results held even when the analysis was restricted to content posted before first depression diagnosis. State-space temporal analysis suggests that onset of depression may be detectable from Twitter data several months prior to diagnosis. Predictive results were replicated with a separate sample of individuals diagnosed with PTSD ($N_{users}$=174, $N_{tweets}$=243,775). A state-space time series model revealed indicators of PTSD almost immediately post-trauma, often many months prior to clinical diagnosis. These methods suggest a data-driven, predictive approach for early screening and detection of mental illness.

**Corresponding authors**: Andrew G. Reece, 33 Kirkland St, Cambridge, MA 02138, 617.395.6841, reece@g.harvard.edu; Christopher M. Danforth, 210 Colchester Ave, Burlington, VT, 05405, chris.danforth@uvm.edu

[a]Department of Psychology, Harvard University, Cambridge, MA 02138; [b]Computational Story Lab, Vermont Advanced Computing Core, and the Department of Mathematics and Statistics, University of Vermont, Burlington, VT 05401; [c]Vermont Complex Systems Center, University of Vermont, Burlington, VT 05401; [d]Department of Management Science and Engineering, Stanford University, Palo Alto, CA 94305; [*]Corresponding authors

Social media data provide valuable clues about physical and mental health conditions. This holds true even in cases where social media users are not yet aware that their health has changed. For example, searching for information on certain health symptoms has been shown to provide accurate early-warning indicators for hard-to-detect cancers (1). Social media networks have been used to plot the trajectory of disease outbreaks (2-4), and to track regional dietary health (5). In addition to physical ailments, predictive screening methods have successfully identified markers in social media data for a number of mental health issues, including addiction (6), depression (7-12), Post-Traumatic Stress Disorder (PTSD) (13,14), and suicidal ideation (15). The field of predictive health screening with social media data is still in its infancy, however, and considerable refinements are needed to develop methodologies that can effectively augment health care. In this report, we present a set of improved methods and novel contributions for predicting and tracking depression and PTSD on Twitter.

Depression has emerged as the leading mental health condition of interest among computational social scientists (7-12), as it is a relatively common mental disorder (16) and influences a range of behaviors and patterns of communication (17). Underdiagnosis of depression remains a persistent problem; a recent survey of a major metropolitan area found nearly half (45%) of all cases of major depression were undiagnosed (18). PTSD, while less common (19), is frequently comorbid with major depression (20). Studies have found that PTSD is underdiagnosed or under-treated by a majority of primary-care physicians (21,22). The costs of underdiagnosis of these conditions, both to human quality of life and health care systems, are considerable. Computational methods for early screening and diagnosis of depression and PTSD have the potential to make a positive impact on a major public health issue, with minimal associated costs and labor intensivity.

**Improvements and novel contributions**

Early efforts to detect depression and PTSD signals in Twitter data have been promising. Park et al. (11) established that Twitter users suffering from depression tended to post tweets containing more negative emotional sentiment compared to healthy users. De Choudhury et al. (7) successfully identified new mothers suffering from postpartum depression, based on changes in Twitter usage and tweet content. In a separate analysis, De Choudhury et al. (8) found that depressive signals were observable in tweets made by individuals with Major Depressive Disorder. A small number of studies have attempted to identify PTSD markers in Twitter data (13,14).

This growing literature has employed progressively more sophisticated methods for making intelligent inferences about Twitter users' mental health based on their online activity. Despite these advances, we identified a number of methodological issues in recent reports which we have improved upon in the present work. A brief review of these modifications provide motivation for the results that follow.

De Choudhury et al. (8) built a predictive model using tweets from depressed individuals posted within a year prior to their self-reported onset of a recent depressive episode[1]. Subjects were included for analysis if they had experienced at least two depressive episodes within that one year period, meaning that the data used to make predictions contained tweets posted after the first onset of depression. The date of first depression diagnosis for each individual was not explicitly accounted for in their model. As a result, model training data may have contained both tweets posted during a previous depressive episode, as well as tweets posted after subjects had already received a formal diagnosis. Both of these possibilities seem especially likely, as depression-related terms such as *diagnosis*, *antidepressants*, *psychotherapy*, and *hospitalization* were significant predictors in their model, along with the names of specific antidepressant medications (eg. *serotonin*, *maprotiline*, and *nefazodone*).

We chose to use only tweets posted prior to the date of subjects' first depression diagnosis, rather than focus on recent depressive episodes, for three reasons. First, self-reported information about depressive symptoms is often inaccurate (23). By contrast, a clinical diagnosis is an explicit event that does not rely on subjective impressions, as may be the case with self-reported onset dates. Second, individuals diagnosed with depression often come to identify with their diagnosis (24,25), and subsequent choices, including how to portray oneself on social media, may be influenced by this identification. It is possible that the predictive signals indicated in De Choudhury et al. (8) were not tracking depressive symptoms, per se, but rather identified purposeful communication choices on the part of depressed Twitter users. Third, and most important, if we are able to accurately discriminate between depressed and healthy participants using only tweets posted prior to first diagnosis, this would support a stronger claim than has been made previously - namely, that Twitter data are capable not only of detecting depression, but can do so before the first diagnosis has been made.

While date of first diagnosis provides a more reliable temporal marker than self-reported onset of symptoms, onset timing is also valuable to researchers and health care professionals looking to better understand depression. This is especially true regarding the onset of an individual's *first* depressive episode. Winokur (26) found that over 50% of depression patients experienced first onset at least 6 months prior to diagnosis. The months during which individuals suffering from depression are undiagnosed and untreated pose a significant health risk. Given that the changes that occur with the first onset of depression may be reflected in social media data, we hypothesized that a computational approach could model the progression of depression without any explicit estimates of onset. Using only the content of participants' tweets, we generated a time series model which charts the course of illness in depressed individuals, and compared this with healthy participants' data.

---

[1] To our knowledge, De Choudhury et al. (8) represent the state-of-the-art in depression screening on Twitter, and our own work was strongly informed by their innovative methods. Accordingly, we report model accuracy scores from De Choudhury et al. (8) along with our results as a point of comparison.

All of the above methodological improvements were also applied to our PTSD analysis, which used a separate cohort of study participants. Extant literature on PTSD detection in Twitter data (13,14) differ from our analysis in important ways. Previous research used bulk collections of public tweets, and assigned PTSD labels to users based on tweets which mentioned a PTSD diagnosis. By comparison, we communicated directly with participants, and excluded any who could not report a specific date on which they received a professional clinical diagnosis. Our analytical approach incorporated a wide array of metadata features and semantic measures, which were limited (14) or missing entirely (13) from earlier research. Most importantly, previous research focused only on differentiating PTSD users from healthy users, without any consideration of timing with respect to the dates of traumatic events or diagnoses. Our models focused specifically on identifying predictive markers of PTSD prior to diagnosis date, as well as tracking the course of this disorder over time.

**Comparison with trained healthcare professionals**

Mitchell, Vaze, and Rao (27) evaluated general practitioners' abilities to correctly diagnose depression in their patients, without assistance from scales, questionnaires, or other measurement instruments. Out of 50,371 patient outcomes culled from 118 studies, 21.9% of patients were actually depressed. General practitioners were able to correctly rule out depression in 81% of non-depressed patients, but only correctly diagnosed depressed patients 42% of the time. Taubman-Ben-Ari et al. (22) tested primary-care physicians' abilities to detect PTSD. PTSD prevalence was 7.5% for men and 10.5% for women in the observed sample (N=683). Physicians correctly identified 2.5% of PTSD cases, and out of all PTSD diagnoses made, only 43% were accurate. We refer to general practitioner accuracy rates (22,27) as an informal benchmark for the quality of our computational model of depression.

# Method

**Data Collection**

Participants were recruited using Amazon's Mechanical Turk (MTurk) crowdwork platform, and we collected user data from both the survey on MTurk and participants' Twitter history[2]. Recruitment and data collection procedures were identical for depression and PTSD samples, with the exception of the condition-specific questionnaire used for screening. Separate surveys were created for affected and healthy samples[3]. In the affected sample surveys, participants were invited to complete a questionnaire that involved passing a series of inclusion criteria, responding to a standardized clinical assessment survey, answering questions related to demographics and mental health history, and sharing social media history. We used the CES-D (Center for Epidemiologic Studies Depression Scale) questionnaire to screen participant depression levels (28). CES-D assessment quality has been demonstrated as on-par with other depression inventories, including the Beck Depression Inventory and the Kellner Symptom Questionnaire (29,30). The Trauma Screening Questionnaire (TSQ) was used to screen for PTSD (31). A comparison cohort of healthy participants were screened to ensure no history of depression or PTSD, respectively, and for active Twitter use. See Appendix I for procedural details.

---

[2] The methods used in recruitment, data collection, and analysis are adopted from Reece and Danforth (12). See Appendix I for procedural details of recruitment and data collection.
[3] The term "affected" here refers to participants affected by either of the mental health conditions of interest: depression or PTSD.

Qualified participants were asked to share their Twitter usernames and history. An app embedded in the survey allowed participants to securely log into their Twitter accounts and agree to share their data. Upon securing consent, we made a one-time collection of participants' Twitter posting history[4]. In total we collected 279,951 tweets from 204 Twitter users for the depression analysis, and 243,775 tweets from 174 Twitter users for the PTSD analysis. Details on participant data protection measures are outlined below.

**Inclusion criteria**

The surveys for affected samples collected age data from participants, and asked qualified participants questions related to their first clinical diagnosis of either depression or PTSD, as well as questions about social media usage at the time of diagnosis. These questions were given in addition to the CES-D or TSQ scales. The purpose of these questions was to determine:
- The date of first clinical diagnosis of the condition,
- Whether or not the individual suspected having the condition before diagnosis, and,
- If so, the number of days prior to diagnosis that this suspicion began

In the case that participants could not recall exact dates, they were instructed to approximate the actual date.

The survey for healthy participants collected age and gender data from participants. It also asked four questions regarding personal health history, which were used as inclusion criteria for this and three other studies. These questions were as follows:
- Have you ever been pregnant?
- Have you ever been clinically diagnosed with depression?
- Have you ever been clinically diagnosed with Post-Traumatic Stress Disorder?
- Have you ever been diagnosed with cancer?

Participants' responses to these questions were not used in analysis, and only served to include qualified respondents in each of the various studies, including the depression- and PTSD-related studies reported here.

**Participant safety and privacy**

---

[4] Twitter allows collection of the most recent 3,200 tweets posted by each user.

This study design raised two important issues regarding ethical research practices, as it concerned both individuals with mental illness and potentially personally identifiable information. We were unable to guarantee strict anonymity to participants, given that usernames and personal information posted to Twitter are often inherently specific to participants' identities (such as usernames and tweets containing real names). As we potentially had the capacity to link study participants' identities to sensitive health information, study participants were informed of the risks of being personally identified from their social media data. Participants were assured that no personal identifiers, including usernames, would ever be made public or published in any format. We used Turk Prime, an interface for conducting MTurk studies, to mask participants' MTurk worker IDs from our records. We made it clear that any links between social media data and private personal health data would be available only to our team of researchers, and participants were able to request to have their data removed at any time.

**Improving data quality**

In an effort to minimize noisy and unreliable data, we applied several quality assurance measures in our data collection process. MTurk workers who have completed at least 100 tasks, with a minimum 95% approval rating, have been found to provide reliable, valid survey responses (32). We restricted survey visibility only to workers with these qualifications. Survey access was also restricted to U.S. IP addresses, as MTurk data collected from outside the United States are generally of poorer quality (33). All participants were only permitted to take the survey once.

We excluded participants with a total of fewer than five Twitter posts. We also excluded participants with CES-D scores of 21 or lower (depression), or TSQ scores of 5 or lower (PTSD). Studies have indicated that a CES-D score of 22 represents an optimal cutoff for identifying clinically relevant depression (34,35); an equivalent TSQ cutoff of 6 has been found to be optimal in the case of PTSD (31).

**Summary statistics**

All data collection took place between February 1, 2016 and June 10, 2016. Across both depressed and healthy groups, we collected data from 204 Twitter users, totaling 279,951 tweets[5]. The mean number of posts per user was 1372.71 (SD=1281.74). This distribution was skewed by a smaller number of frequent posters, as evidenced by a median value of just 861 posts per user. See Table 1 for summary statistics.

In the depressed group, 147 crowdworkers successfully completed participation and provided access to their Twitter data. Imposing the CES-D cutoff reduced the number of viable participants to 105. The mean age for viable participants was 30.3 years (SD=8.34), with a range of 18 to 64 years. Dates of participants' first depression diagnoses ranged from March 2010 to

---

[5] This number includes up to 3,200 tweets from each participant's Twitter history. The analyses in this report focus only on tweets from depressed users created before the date of first depression diagnosis.

February 2016, with nearly all diagnosis dates (92%) occurring in the period 2013-2015. In the healthy group, 99 participants completed participation and provided access to their Twitter data. The mean age for this group was 33.9 years, with a range of 19 to 63 years, and 42% of respondents were female. (Gender data were not collected for affected sample surveys.)

For the PTSD analysis, we collected data from 174 Twitter users, totaling 243,775. The mean number of posts per user was 1372.71 (SD=1281.74). This distribution was skewed by a smaller number of frequent posters, as evidenced by a median value of just 862 posts per user.

In the PTSD sample, 73 crowdworkers successfully completed participation and provided access to their Twitter data. Imposing the TSQ cutoff reduced the number of viable participants to 63. The mean age for viable participants was 30.64 years (SD=7.57), with a range of 21 to 54 years. Dates of participants' first PTSD diagnoses ranged from April 2010 to December 2015, with nearly all diagnosis dates (94%) occurring in the period 2013-2015. In the healthy group, 111 participants completed participation and provided access to their Twitter data. The mean age for this group was 33.25 years, with a range of 19 to 63 years, and 51% of respondents were female.

| Depression | Users | Posts | Posts $\mu(\sigma)$ | Posts (median) |
|---|---|---|---|---|
| Total | 204 | 279,951 | 1373 (1282) | 862 |
| Depressed | 105 | 164,218 | 1564 (1332) | 1127 |
| Healthy | 99 | 115,733 | 1169 (1200) | 574 |
| PTSD | Users | Posts | Posts $\mu(\sigma)$ | Posts (median) |
| Total | 174 | 243,775 | 1401 (1284) | 946.5 |
| Depressed | 63 | 91,589 | 1564 (1332) | 1058 |
| Healthy | 111 | 152,186 | 1371 (1268) | 893 |

Table 1. Summary statistics for depression and PTSD tweet collection ($N_{depr}$=279,951, $N_{ptsd}$=243,775).

**Feature extraction**

We extracted several categories of predictors from the Twitter posts collected. Predictor selection for both depression and PTSD was based on prior machine learning models of depression in Twitter data (7,8), as the two conditions' high comorbidity rates suggest their predictive signals may exhibit considerable overlap (20). Depressed Twitter users have been observed to tweet less frequently than non-depressed users (8), so we used total tweets per user, per day, as a measure of user activity. Tweet metadata was analyzed to assess average word count per tweet[6], whether or not the tweet was a retweet, and whether or not the tweet was a reply to someone else's tweet. The labMT, LIWC 2007, and ANEW unigram sentiment instruments were used to quantify the happiness of tweet language (36-39). The use of labMT, which has shown strong prior performance in analyzing happiness on Twitter (40, 41), is novel with respect to depression screening; ANEW and LIWC have been successfully applied in previous studies on depression and Twitter (7,8,14). LIWC was also used to compile frequency counts of various parts of speech (e.g., pronouns, verbs, adjectives) and semantic categories (e.g., food words, familial terms, profanity) as additional predictors (36).

---
[6] A word is defined as a set of characters surrounded by whitespace.

**Units of observation**

Determining the best time span for analysis raises a difficult question: When and for how long does mental illness occur? Receiving a clinical diagnosis of depression or PTSD does not imply that an individual remains in a persistent state of illness, and so to conduct analysis with an individual's entire posting history as a single unit of observation is a dubious proposition. At the other extreme, to take one tweet as a unit of observation runs the risk of being too granular.

DeChoudhury et al. (8) looked at all of a given user's tweets in a single day, and aggregated those data into per-person, per-day units of observation. In this report we have followed the convention of aggregated "user-days" as a primary unit of analysis, rather than try to categorize a person's entire history, or analyze each individual tweet[7]. In our own previous research, however, we have found that many Twitter users do not generate enough daily content to make for robust unigram sentiment analysis (42). For completeness, we conducted analyses using both daily and weekly units of observation[8]. Both analyses yielded predictive models of similar strengths, with the weekly model showing a slight, but consistent, edge in performance. We report accuracy metrics from both analyses, but restrict other results to the daily-unit analysis to allow for more direct comparison with previous research. Details of weekly-unit analytical results are available in Appendix II.

**Statistical framework**

*Machine learning models*

We trained supervised machine learning classifiers to discriminate between affected and healthy sample members' observations. Classifiers were trained on a randomly-selected 70% of total observations, and tested on the remaining 30%. Out of several candidate algorithms, a 1200-tree Random Forests classifier demonstrated best performance. Stratified five-fold cross-validation was used to to optimize Random Forests hyperparameters, and final accuracy scores were averaged over five separate randomized runs. Precision, recall, specificity, negative predictive value, and F1 accuracy scores are reported, and general practitioners' unassisted diagnostic accuracy rates as reported in Mitchell, Vaze, and Rao (27) (MVR) and Taubman-Ben-Ari et al. (22) (TBA) are used as informal benchmarks for depression and PTSD, respectively[9]. See Supplementary Appendix III for optimization details.

---

[7] Occasionally, when reporting results we refer to observations or tweets as "depressed", eg. "depressed tweets received fewer likes". It would be more correct to use the phrase "tweet data from depressed participants, aggregated by user-days" instead of "depressed tweets", but we chose to sacrifice a degree of technical correctness for the sake of clarity.
[8] We also considered hourly units of observation, as De Choudhury et al. (8) found substantial differences in diurnal patterns between depressed and healthy subjects. In our sample, however, exploratory analysis showed no differences in hourly posting trends between these groups, and we did not conduct analyses using hourly units.

*Time series analysis*

Predictive screening methods use indirect indicators, such as language use on social media, to infer health status. We are not actually interested in the average word count of depressed individuals' tweets, for example, but rather we hope that this measure will allow us some access to the underlying variable we truly care about: depression. Accordingly, state-space models, which use observable data to estimate the status of a latent, or hidden, variable over time, may provide useful insights. We trained a two-state Hidden Markov Model (HMM) to detect differential changes between affected and healthy groups over time.

The use of HMM presents an interpretability challenge: how to know whether resulting latent states have any relationship to the clinical condition of interest? Consider the case of depression: Finding evidence that HMM had, in fact, recovered two states from our data that closely resembled the depressed and healthy classes was prerequisite to making any inferences based on HMM output. We addressed this issue by comparing HMM output with mean differences between depressed and healthy observations in the raw data. If the direction of the differences between HMM mean parameter estimates generally agree with the true differences in the data, this provides evidence that the two sample groups in our data (depressed and healthy) are well-characterized by HMM latent states. For example, if depressed observations contained more sad words on average than healthy observations (variable name: "LIWC_sad"), then the HMM state with the higher LIWC_sad estimate is more likely to be the depressed one, given that HMM does track depression (ie. the latent states generated by HMM map onto "depressed" and "healthy"). If, on the other hand, HMM-generated states are weakly or not at all related to depression, there should be no clear alignment between HMM means and means in the raw data. The same procedure was applied when fitting an HMM to the PTSD data.

**Word shift graphs**

Machine learning algorithms provide powerful predictive capability, but most algorithms offer little in the way of context and interpretation. Word shift graphs use the labMT happiness scores, the most important predictor for both depression and PTSD analyses, to show qualitatively how inter-group differences may be driven by the usage of specific words in tweets (38). We present word shift graphs comparing the way tweet language adjusted happiness scores in affected and healthy samples. The visualization ranks words by their contribution to the happiness difference between the two groups (for more explanation of word shift rankings, see 43, 44). Word shifts are generated from a different statistical method than the machine learning algorithm we used to make predictions, and so should be treated as exploratory analyses independent of our main findings.

---

[9] Comparing point estimates of accuracy metrics is not a statistically robust means of model comparison, in addition to the fact that our results are drawn from different samples, using different observational units, than our chosen comparisons. However, we felt it was more meaningful to frame our findings in a realistic context, rather than to benchmark against a naive statistical model that simply predicted the majority class for all observations.

# Results

For the depression study, we analyzed 74,990 daily observations (23,541 depressed) from 204 individuals (105 depressed). For the PTSD study, we analyzed 54,197 daily observations (13,008 PTSD) from 174 individuals (63 PTSD). Observations from affected sample members accounted for 31.4% and 24% of the entire data sets for depression and PTSD, respectively.

**Machine learning classifier**

Results are reported for both daily and weekly units of observation (see Table 2 and Figure 1).

Our best depression classifier, averaged over cross-validation iterations, improved over both Mitchell et al. (27) and De Choudhury et al. (8) on several metrics. Our depression model's precision rate was considerably higher, with just over 1 false positive for every 10 depression diagnoses. By comparison, general practitioners from Mitchell et al. (27) incorrectly diagnosed patients as having depression in more than half of all diagnoses.

Our best PTSD classifier improved considerably over the primary-care physicians from Taubman-Ben-Ari et al. (22) (TBA). Whereas more than half of all PTSD diagnoses made by TBA physicians were incorrect, our model was correct in roughly 9 out of every 10 (88.2%) of its PTSD predictions. Model recall rate was strong, with 68.3% discovery of actual PTSD sample observations.

The labMT happiness score was the strongest predictor of both depression and PTSD. Notably, average labMT average happiness over user days showed only modest correlation with ANEW ($r_{depr}$=.37, $r_{ptsd}$=.36) and LIWC ($r_{depr}$=.36, $r_{ptsd}$=.37), suggesting that labMT identifies relevant prediction signals not fully captured by other sentiment instruments. The additional benefit offered by labMT in this context may be a reflection of its inclusion of the 5000 most frequently used words on Twitter, including slang (38). The second most important variable was word count, which represented the average number of words per tweet. Sentiment-related variables from ANEW and LIWC accounted for most of the remaining top predictors (see Figure 1)[10].

---

[10] As with most decision-tree classifiers, the Random Forests algorithm provides information on the relevance, but not the directionality, of predictors. In other words, we can know how important a variable was to the algorithm, but not if it was positively or negatively associated with the response variable. Word shift graphs, reported below, offer some indication of directionality but are computed differently than Random Forests and should not be used to directly interpret Random Forests output.

| Depression | MVR μ | DC μ | Daily μ (σ) | Weekly μ (σ) |
|---|---|---|---|---|
| Recall | .510 | .614 | .518 (.000) | .521 (.000) |
| Specificity | .813 | N/A | .958 (.000) | .969 (.000) |
| Precision | .42 | .742 | .852 (.000) | .866 (.000) |
| NPV | .858 | N/A | .812 (.000) | .841 (.000) |
| F1 | .461 | .672 | .644 (.000) | .651 (.000) |
| PTSD | TBA μ | NHC μ | Daily μ (σ) | Weekly μ (σ) |
| Recall | .249 | .82 | .683 (.000) | .658 (.000) |
| Specificity | .979 | N/A | .988 (.000) | .994 (.000) |
| Precision | .429 | .86 | .882 (.000) | .934 (.000) |
| NPV | .602 | N/A | .959 (.000) | .954 (.000) |
| F1 | .315 | .84 | .769 (.000) | .772 (.000) |

Table 2. Classification accuracy metrics for daily and weekly models ($N_{depr}$=74,990, $N_{ptsd}$=54,197). Accuracy scores from Mitchell et al. (28) (MVR), De Choudhury et al. (8) (DC), Taubman-Ben-Ari et al. (22) (TBA), and Nadeem, Horn, & Coppersmith (13) (NHC) are included for comparison to depression (MVR, DC) and PTSD (TBA, NHC) results. Table cells marked N/A indicate unavailable metrics from previous studies.

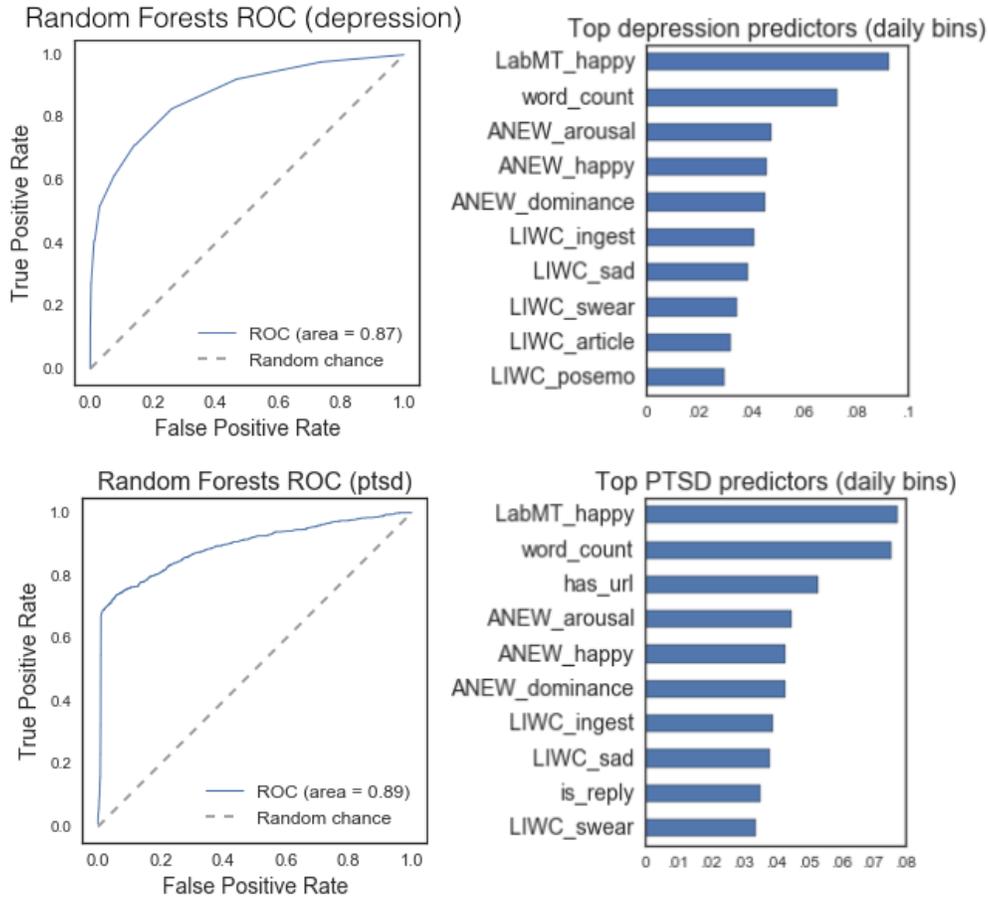

Fig. 1. ROC curve and top predictors for Random Forests algorithm, for depression and PTSD samples ($N_{depr}$=74,990, $N_{ptsd}$=54,197). Predictor names ending in "_happy" are happiness measures; LIWC predictors (36) refer to the occurrence of semantic categories (eg. LIWC_ingest refers to food and eating words, LIWC_swear refers to profanity).

**Time series analysis**

A Hidden Markov Model simulated affected and healthy states. HMM states were determined to accurately track with affected and healthy groups by comparing differences in mean parameter estimates between the model fit and original data. Across all 40 predictors, HMM means were in agreement with true means for the depression sample in 38 cases (95% agreement), and were in 100% agreement with true means for the PTSD sample. This evidence strongly suggested that the two states identified by HMM were closely aligned with the affected and healthy classes in our data, and we have reported HMM results based on this assumption. See Figures 2 and 3.

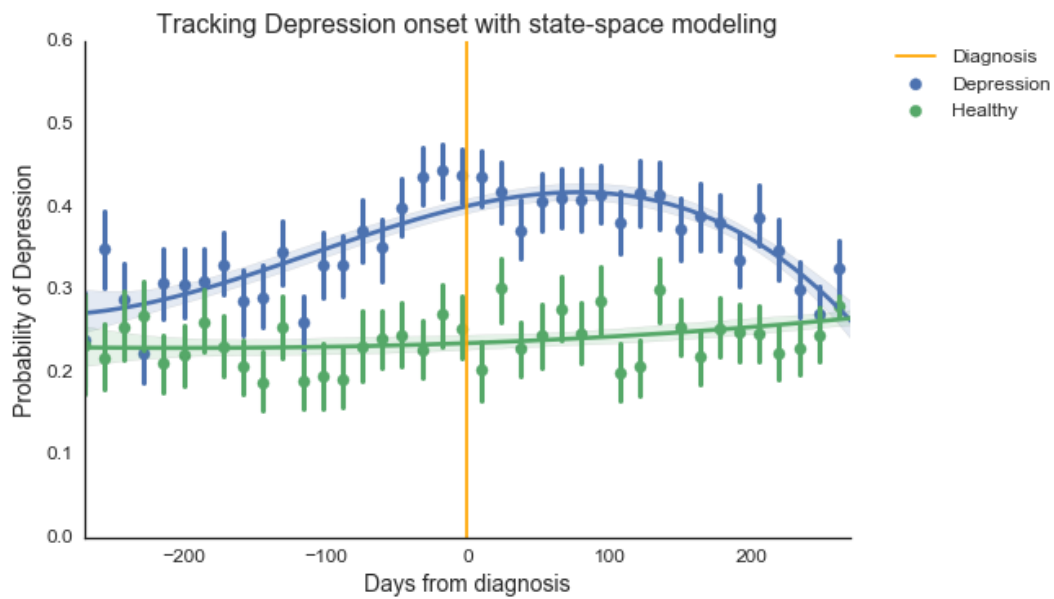

Figure 2. Hidden Markov Model showing probability of depression (N=74,990). X-axis represents days from diagnosis. Healthy data are plotted from a consecutive time span of equivalent length. Trend lines represent cubic polynomial regression fits with 95% CI bands, points are aggregations of 14 day periods, with error bars indicating 95% CI on central tendency of daily values.

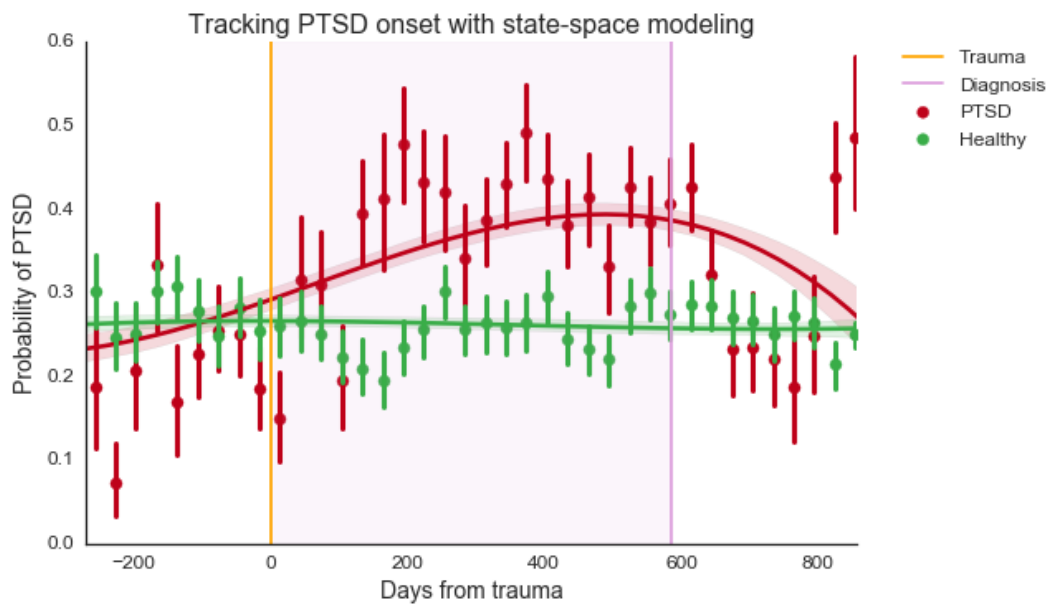

Figure 3. Hidden Markov Model showing probability of PTSD (N=54,197). X-axis represents days from trauma event. Healthy data are plotted from a consecutive time span of equivalent length. The purple vertical line indicates mean number of days to PTSD diagnosis, post-trauma, and the purple shaded region shows the average period between trauma and diagnosis. Trend lines represent cubic polynomial regression fits with 95% CI bands, points are aggregations of 30 day periods, with error bars indicating 95% CI on central tendency of daily values.

Depressed individuals showed a slightly higher probability of depression even at nine months prior to diagnosis, and gradually diverged from healthy data points, becoming pronounced well before diagnosis. Healthy individuals showed a steady, lower probability of depression, which did not change noticeably over an 18-month period. By three months prior to diagnosis, depressed subjects showed a marked rise in probability of being in a depressed state, whereas healthy individuals showed little or no change over the same time period. Post-diagnosis, probability of depression began to decrease after 3-4 months (90-120 days). This trajectory matches closely with average improvement time frames observed in therapeutic programs (45). Given that HMM constructed latent states from unlabeled data, it is striking that HMM not only reconstructed the division between depressed and healthy groups, but also generated a plausible timeline for depression onset and recovery. Similarly, tweets from individuals with PTSD deviated from healthy tweets within weeks after the date of the traumatic event that caused PTSD (indicated by the orange line in Figure 3), and well over a year before the average time to diagnosis (the mean time period from trauma to diagnosis was 586 days). A decrease in PTSD probability can be observed occurring shortly after diagnosis, indicating possible improvement due to treatment.

**Word shift graphs**

We averaged labMT happiness scores across observations in each class, after the removal of common neutral words and re-tweeted promotional material (38)[11,12]. We observed that tweets authored by the depressed class were sadder ($h_{avg}$ = 6.01) than the healthy class ($h_{avg}$ = 6.15). In Figure 4, we rank order individual words with respect to their contribution to this observed difference, and display the top contributing words. PTSD word shift graphs are included in Appendix IV.

The dominant contributor to the difference between depressed and healthy classes was an increase in usage of negative words by the depressed class, including "don't", "no", "not", "murder", "death", "never", and "sad". The second largest contributor was a decrease in positive language by the depressed class, relative to the healthy class, including fewer appearances of "photo", "happy", "love", and "fun". The increased usage of negatively-valenced language by depressed individuals is congruent with previous research (46).

---

[11] Words were removed with labMT happiness scores between 4 and 6, on a 1-9 scale. This includes many common parts of speech, including articles and pronouns, which contribute little to understanding inter-group differences in valenced language.

[12] Some of the positive language observed more frequently among healthy individuals came from re-tweets of promotional or other advertising material (e.g., "win", "free", "gift"). We removed obvious promotional retweets when generating word shift graphs, as their removal did not significantly change mean tweet-happiness differences between groups, and the resulting graphs gave better impressions of what participants personally tweeted about.

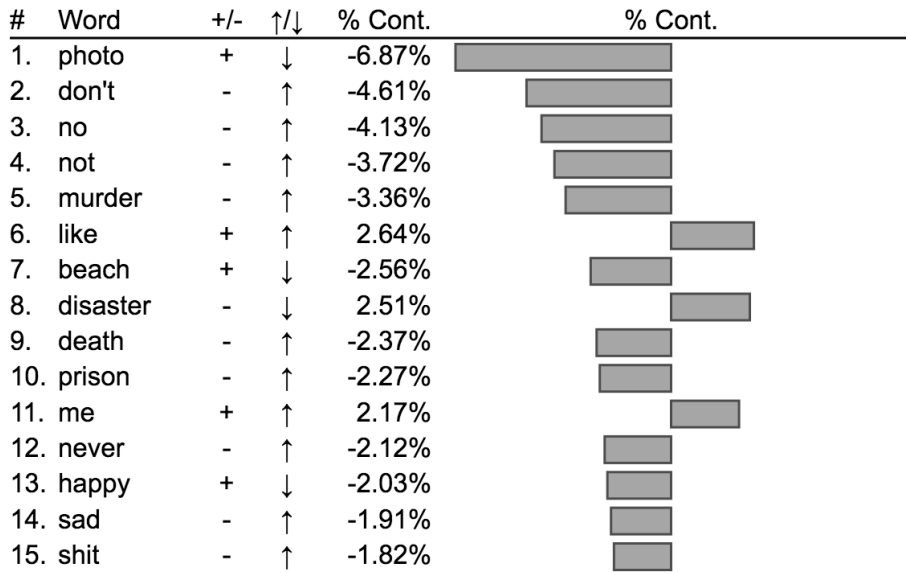

Figure 4. Depression word-shift graph revealing contributions to difference in Twitter happiness observed between depressed and healthy participants. In column 3, (-) indicates a relatively negative word, and (+) indicates a relatively positive word, both with respect to the average happiness of all healthy tweets. An up (down) arrow indicates that word was used more (less) by the depressed class. Words on the left (right) contribute to a decrease (increase) in happiness in the depressed class. See Appendix IV for PTSD word-shift graph.

## Discussion

The aim of the present study was to identify predictive markers of depression and PTSD based on users' Twitter data using computational methods. Our findings strongly support the claim that computational methods can effectively screen Twitter data for indicators of depression and PTSD. Our method identified these mental health conditions earlier and more accurately than the typical performance of trained health professionals, and was more precise than previous computational approaches. Our state-space models portrayed a timeline for depression which is impressively realistic, given that it was generated analyzing only the text of 140-character messages. In addition, HMM identified a rise in probability of PTSD within six months, post-trauma, compared to the average 19 month delay between trauma event and diagnosis experienced by the individuals in our sample. Word shifts provided context for the specific differences in language that shifted happiness scores between samples. These advances make improvements on existing predictive screening technology, as well as contribute novel methods for the identification and tracking of mental illness.

The HMM depression timeline is an intriguing finding, and should be treated with both optimism and caution. HMM assigned each data point a probability of belonging to two latent state-spaces while "blind" to our actual states of interest, as affected/healthy labels were removed from HMM training data. Considering that the model could have used criteria completely unrelated to mental health to delineate between the two latent classes, it is noteworthy that the resulting states' mean estimates for each variable in both PTSD and depression analyses closely resembled the mean estimates for affected and healthy participants' data, respectively. This adds support for the claim that affected-condition and healthy Twitter users' data are objectively different, in addition to providing justification for the use of HMM assignments as indicators of depression/PTSD signals at a given point in time. Despite this evidence in favor of applying HMM to analyze mental health trajectories, HMM is an unsupervised learning procedure and so conclusive HMM-based inferences should be approached cautiously and with close attention on validation procedures. The diverging trajectories observed in our HMM time series suggest that, with careful attention to model validity, state-space modeling may be used to identify and track the onset of certain mental illnesses over time, using only Twitter data.

The labMT happiness measure proved to be the most important predictor in our model, and was considerably stronger than ANEW or LIWC happiness indicators. This is in line with a series of previous findings, which have found labMT measures to be a superior for tracking happiness in Twitter data (39), and suggests that future research in this field should incorporate this instrument for more accurate measurements. That average tweet word count was the second most important predictor is intriguing, especially as increases in word count were positively associated with depression and PTSD. If anything, depression is often characterized by reduced communication (17), although word count is distinct from posting frequency, which was not a significant predictor in our models[13]. The current depression and PTSD literatures are largely devoid of studies relating verbosity to these conditions, and so this finding may motivate new inquiries into behavioral traits of mental health disorders as observed on social media.

From a practical standpoint, our model showed considerable improvement over the ability of unassisted general practitioners to correctly diagnose depression and PTSD. Despite the imprecise nature of this comparison, given the paucity of data currently available to serve as benchmarks for the type of analysis performed in the present study, our model's relative success seems encouraging. Health care providers may be able to improve quality of care and better identify individuals in need of treatment based on the simple, low-cost methods outlined in this report. Especially given that mental health services are unavailable or underfunded in many countries (47), this computational approach, which only requires patients' digital consent to share their social media history, may open avenues to care which are currently difficult or impossible to provide. Future investigations would use the same participant pool to collect both health professionals' assessments, as well as computational models of participant social media behavior, to allow for more precise comparison.

The present findings may be limited by the non-specific use of the term "depression" in participant surveys. While earlier research identified depression predictors in Twitter data for Major Depressive Disorder and postpartum depression (7,8), we used a more general category in our recruitment and data collection to build a predictive model capable of screening for common depressive signals. We acknowledge that depression diagnoses exist across a clinical spectrum. It is possible that participants with a specific type of depression were responsible for the observed results. Future research might examine other specific depression classes, including manic depression and dysthymia, to determine whether predictive screening models should be segmented per diagnosis type.

---

[13] In a separate study of Instagram users, we did observe a positive correspondence between depression and posting frequency (12).

It is also possible that inferences from our results are limited specifically to Twitter users who have been diagnosed with depression or PTSD, and who are willing to share their social media history with researchers. Current literature on depression treatment suggests that people who seek out mental health services are usually "well-informed and psychologically minded, experience typical symptoms of depression and little stigma, and have confidence in the effectiveness of treatment, few concerns about side effects, adequate social support, and high self-efficacy" (48). Since it is possible only a subset of Twitter users will fit this description, we recommend making conservative inferences about depression, as well as PTSD, based on our findings.

Considering the frequent comorbidity of depression and PTSD (20), together with the similarity in predictor importance observed across our analyses of these two conditions, the signals driving our predictive models may share considerable overlap. While our results do not offer strict segregation between these two conditions, this gives little cause for concern when considered from a mental health screening perspective. If the desired outcome is to identify individuals who may be in need of mental health services, whether an individual is flagged for evaluation for depression with possible associated PTSD, or vice-versa, becomes an academic distinction. If anything, the issue of comorbidity may serve as a useful reminder that this computational method should not be regarded as a standalone diagnostic tool, but rather as a technology for early identification of potential mental health issues.

As the methods employed in the present study aim to infer health related information about individuals, some additional cautionary considerations are in order. Data privacy and ethical research practices are of particular concern, given recent admissions that individuals' Facebook and dating profile data were experimentally manipulated or exposed without permission (49,50). Indeed, we observed a response rate reflecting a seemingly reluctant population. Of the 2,261 individuals who began our survey, 790 (35%) refused to share their Twitter username and history, even after we identified ourselves as an "academic, not-for-profit research team" and provided the above-mentioned guarantees about data privacy. Future research should prioritize establishing confidence among experimental participants that their data will remain secure and private. Complicating efforts to build socio-technical tools such as the models presented in this study, data trends often change over time, degrading model performance without frequent calibration (51). As such, our results should be considered a methodological proof-of-concept upon which to build and refine subsequent models.

This report provides an outline for an accessible, accurate, and inexpensive means of improving depression and PTSD screening, especially in contexts where in-person assessments are difficult or costly. In concert with robust data privacy and ethical analytics practices, future models based on our work may serve to augment traditional mental health care procedures. More generally, our results support the idea that computational analysis of social media can be used to identify major changes in individual psychology.

# References

1. Paparrizos J, White RW, Horvitz E (2016) Screening for Pancreatic Adenocarcinoma Using Signals From Web Search Logs: Feasibility Study and Results. *Journal of Oncology Practice*: JOPR010504. doi.org/10.1200/JOP.2015.010504
2. Christakis NA, Fowler JH (2010) Social network sensors for early detection of contagious outbreaks. *PLoS ONE 5*(9): e12948. doi.org/10.1371/journal.pone.0012948
3. Li J, Cardie C (2013) Early Stage Influenza Detection from Twitter. *arXiv:1309.7340 [cs]*. Available at http://arxiv.org/abs/1309.7340.
4. Schmidt, CW (2012). Trending now: Using social media to predict and track disease outbreaks. *Environ Health Perspect 120*(1): a30–a33. doi.org/10.1289/ehp.120-a30.
5. Alajajian SE, Williams JR, Reagan AJ, Alajajian, SC, Frank, MR et al. (2015) The Lexicocalorimeter: Gauging public health through caloric input and output on social media. *arXiv preprint arXiv:1507.05098*.
6. Moreno M, Christakis D, Egan K, Brockman L, Becker T (2012) Associations between displayed alcohol references on facebook and problem drinking among college students. *Arch Pediatr Adolesc Med 166*(2): 157–163. doi.org/10.1001/archpediatrics.2011.180
7. De Choudhury M, Counts S, Horvitz E (2013) Predicting postpartum changes in emotion and behavior via social media. In *Proceedings of the SIGCHI Conference on Human Factors in Computing Systems* (ACM: New York), pp. 3267–3276. doi.org/10.1145/2470654.2466447
8. De Choudhury M, Gamon M, Counts S, Horvitz E (2013) Predicting depression via social media. In *Seventh International AAAI Conference on Weblogs and Social Media*.
9. Katikalapudi R, Chellappan S, Montgomery F, Wunsch D, Lutzen K (2012) Associating internet usage with depressive behavior among college students. *IEEE Tech Soc Magazine 31*(4): 73–80. doi.org/10.1109/MTS.2012.2225462
10. Moreno MA, Jelenchick LA, Egan KG, Cox E, Young H, Gannon KE, Becker T (2011) Feeling bad on Facebook: depression disclosures by college students on a social networking site. *Depress Anxiety 28*(6): 447–455. doi.org/10.1002/da.20805
11. Park G, Schwartz HA, Eichstaedt JC, Kern ML, Kosinski M, Stillwell, DJ et al. (2015) Automatic personality assessment through social media language. *J Pers Soc Psychol 108*(6): 934–952. http://doi.org/10.1037/pspp0000020
12. Reece, AG, Danforth, CM (2016) Instagram photos reveal predictive markers of depression. http://arxiv.org/abs/1608.03282
13. Nadeem M, Horn M, Coppersmith G (2016) Identifying depression on Twitter. *arXiv:1607.07384 [cs, Stat]*. Available at http://arxiv.org/abs/1607.07384. Accessed on August 13 2016.
14. Coppersmith G, Harman C, Dredze M (2014) Measuring post traumatic stress disorder in Twitter. In *Eighth International AAAI Conference on Weblogs and Social Media*.
15. De Choudhury M, Kiciman E, Dredze M, Coppersmith G, Kumar M (2016) Discovering shifts to suicidal ideation from mental health content in social media. In *Proceedings of the 2016 CHI Conference on Human Factors in Computing Systems* (ACM: New York), pp. 2098–2110. doi.org/10.1145/2858036.2858207
16. Ferrari A, Somerville AJ, Baxter AJ, Norman R, Patten SB, Vos T, Whiteford HA (2013) Global variation in the prevalence and incidence of major depressive disorder: a systematic review of the epidemiological literature. *Psychol Med 43*(3): 471–481. doi.org/10.1017/S0033291712001511
17. American Psychiatric Association (2000) Diagnostic and statistical manual of mental disorders (4th ed., text rev.). doi:10.1176/appi.books.9780890423349.
18. Gwynn RC, McQuistion HL, McVeigh KH, Garg RK, Frieden TR, Thorpe LE (2008) Prevalence, diagnosis, and treatment of depression and generalized anxiety disorder in a diverse urban community. *Psychiatr Serv 59*(6): 641–647. doi.org/10.1176/ps.2008.59.6.641

19. Stein MB, McQuaid JR, Pedrelli P, Lenox R, McCahill ME (2000) Posttraumatic stress disorder in the primary care medical setting. *Gen Hosp Psychiatry 22*(4): 261–269. doi.org/10.1016/S0163-8343(00)00080-3
20. Campbell DG, Felker BL, Liu CF, Yano EM., Kirchner JE, Chan D et al.. (2007) Prevalence of Depression–PTSD comorbidity: Implications for clinical practice guidelines and primary care-based interventions. *J Gen Intern Med 22*(6): 711–718. doi.org/10.1007/s11606-006-0101-4
21. Munro CG, Freeman CP, Law R (2004) General practitioners' knowledge of post-traumatic stress disorder: a controlled study. *Br J Gen Pract 54*(508): 843–847.
22. Taubman-Ben-Ari, O, Rabinowitz J, Feldman D, Vaturi R (2001) Post-traumatic stress disorder in primary-care settings: prevalence and physicians' detection. *Psychol Med*, *31*(03): 555–560. doi.org/10.1017/S0033291701003658
23. Eaton WW, Neufeld K, Chen L, Cai G (2000) A comparison of self-report and clinical diagnostic interviews for depression: Diagnostic interview schedule and schedules for clinical assessment in neuropsychiatry in the baltimore epidemiologic catchment area follow-up. *Arch Gen Psychiatry 57*(3): 217–222.doi.org/10.1001/archpsyc.57.3.217
24. Cornford CS, Hill A, Reilly J (2007) How patients with depressive symptoms view their condition: a qualitative study. *Fam Pract 24*(4): 358–364. doi.org/10.1093/fampra/cmm032
25. Karp DA (1994) Living with depression: Illness and identity turning points. *Qual Health Res 4*(1): 6–30. doi.org/10.1177/104973239400400102
26. Winokur G (1976) Duration of Illness prior to Hospitalization (Onset) in the Affective Disorders. *Neuropsychobiology*, *2*(2-3), 87–93.doi.org/10.1159/000117535
27. Mitchell AJ, Vaze A, Rao S (2009) Clinical diagnosis of depression in primary care: a meta-analysis. *Lancet 374*(9690): 609–619. doi.org/10.1016/S0140-6736(09)60879-5
28. Radloff LS (1977) The CES-D scale: A self-report depression scale for research in the general population. *Appl Psych Manage 1*(3): 385–401. doi.org/10.1177/014662167700100306
29. Fountoulakis KN, Bech P, Panagiotidis P, Siamouli M, Kantartzis S, Papadopoulou A et al. (2007) Comparison of depressive indices: Reliability, validity, relationship to anxiety and personality and the role of age and life events. *J Affect Disord 97*(1–3): 187–195. doi.org/10.1016/j.jad.2006.06.015
30. Zich JM, Attkisson CC, Greenfield TK (1990) Screening for depression in primary care clinics: The CES-D and the BDI. *Int J Psychiatry Med 20*(3): 259–277. doi.org/10.2190/LYKR-7VHP-YJEM-MKM2
31. Brewin CR, Rose S, Andrews B, Green J, Tata P, McEvedy YC et al. (2002) Brief screening instrument for post-traumatic stress disorder. *Br J Psychiatry 181*(2): 158–162.doi.org/10.1192/bjp.181.2.158
32. Peer E, Vosgerau J, Acquisti A (2013) Reputation as a sufficient condition for data quality on Amazon Mechanical Turk. *Behav Res Methods 46*(4): 1023–1031. doi.org/10.3758/s13428-013-0434-y
33. Litman L, Robinson J, Rosenzweig C (2014) The relationship between motivation, monetary compensation, and data quality among US- and India-based workers on Mechanical Turk. *Behav Res Methods 47*(2): 519–528. doi.org/10.3758/s13428-014-0483-x
34. Cuijpers P, Boluijt P, van Straten A (2007) Screening of depression in adolescents through the Internet. *Eur Child Adolesc Psychiatry 17*(1): 32–38. doi.org/10.1007/s00787-007-0631-2
35. Haringsma R, Engels GI, Beekman ATF, Spinhoven P (2004) The criterion validity of the Center for Epidemiological Studies Depression Scale (CES-D) in a sample of self-referred elders with depressive symptomatology. *Int J Geriatr Psychiatry 19*(6): 558–563. doi.org/10.1002/gps.1130
36. Pennebaker JW, Boyd RL, Jordan K, Blackburn K (2015) The development and psychometric properties of LIWC2015. *UT Faculty/Researcher Works*.
37. Bradley MM, Lang PJ (1999) Affective norms for English words (ANEW): Instruction manual and affective ratings. Technical report C-1, the center for research in psychophysiology, University of Florida.

38. Dodds PS, Harris KD, Kloumann IM, Bliss CA, Danforth CM (2011) Temporal Patterns of Happiness and Information in a Global Social Network: Hedonometrics and Twitter. *PLoS ONE*, *6*(12), e26752. doi.org/10.1371/journal.pone.0026752
39. Reagan AJ, Tivnan BF, Williams JR, Danforth CM, Dodds PS (2016) Benchmarking sentiment analysis methods for large-scale texts: A case for using continuum-scored words and word shift graphs.http://arxiv.org/abs/1512.00531
40. Cody EM, Reagan AJ, Mitchell L, Dodds PS, Danforth CM (2015) Climate change sentiment on Twitter: An unsolicited public opinion poll. *PLoS one 10*(8): e0136092.
41. Frank MR, Mitchell L, Dodds PS, Danforth CM (2013) Happiness and the patterns of life: A study of geolocated tweets. *Scientific Reports 3(2625)*.
42. Bliss CA, Kloumann IM, Harris KD, Danforth CM, Dodds PS (2012) Twitter reciprocal reply networks exhibit assortativity with respect to happiness. *Journal of Computational Science 3(5)*: pp. 388-397.
43. Dodds PS, Clark EM, Desu S, Frank MR, Reagan AJ, Williams JR et al. (2015) Human language reveals a universal positivity bias. *Proceedings of the National Academy of Sciences*, *112*(8), 2389–2394. doi.org/10.1073/pnas.1411678112
44. Storylab (2014) Hedonometer 2.0: Measuring happiness and using word shifts. Available at http://www.uvm.edu/storylab/2014/10/06/hedonometer-2-0-measuring-happiness-and-using-word-shifts/. Accessed on August 14, 2016.
45. Rude S, Gortner EM, Pennebaker J (2004) Language use of depressed and depression-vulnerable college students. *Cogn Emotion*,*18*(8): 1121–1133. doi.org/10.1080/02699930441000030
46. Schulberg HC, Katon W, Simon GE, Rush A (1998) Treating major depression in primary care practice: An update of the agency for health care policy and research practice guidelines. *Arch Gen Psychiatry 55*(12): 1121–1127. doi.org/10.1001/archpsyc.55.12.1121
47. Detels R (2009) *The scope and concerns of public health*. Oxford University Press.
48. Epstein RM, Duberstein PR, Feldman MD, Rochlen AB, Bell RA, Kravitz RL, et al (2010) "I didn't know what was wrong:" How people with undiagnosed depression recognize, name and explain their distress. *J Gen Intern Med 25*(9): 954–961. doi.org/10.1007/s11606-010-1367-0
49. Fiske ST, Hauser RM (2014) Protecting human research participants in the age of big data. *Proc Natl Acad Sci USA 111*(38): 13675–13676. doi.org/10.1073/pnas.1414626111
50. Lumb D (2016) Scientists release personal data for 70,000 OkCupid profiles. Available at engt.co/2b4NnQ0. Accessed August 7, 2016.
51. Lazer D, Kennedy R, King G, Vespignani A (2014) The parable of Google Flu: Traps in big data analysis. *Science 343*(6176): 1203–1205. doi.org/10.1126/science.1248506

**Acknowledgments**. A.G.R. was supported by the Sackler Scholar Programme in Psychobiology. P. S. D. and C. M. D. were supported by NSF grant #1447634. We thank L. Doyle for conversations and P. Mair for manuscript review.

**Author Contributions.** A.G.R., A.J.R., and C.M.D. contributed to design. A.G.R., A.J.R., and C.M.D. contributed to analysis. A.G.R., A.J.R., C.M.D., E.J.L., K.L.M.L., and P.S.D. contributed to writing.

# Supplementary Information

I. Research protocol

The present study was reviewed and approved by the Harvard University Institutional Review Board, approval #15-2529 as well as the University of Vermont Institutional Review Board, approval #CHRMS-16-135. All study participants were informed of and acknowledged all of the study goals, expectations, and procedures, including data privacy, prior to any data collection. Surveys were built using the Qualtrics survey platform, and analyses were performed using Python and R. Twitter data collection apps were written in Python, using the Twitter developer's Application Programming Interface (API).

II. Weekly model output

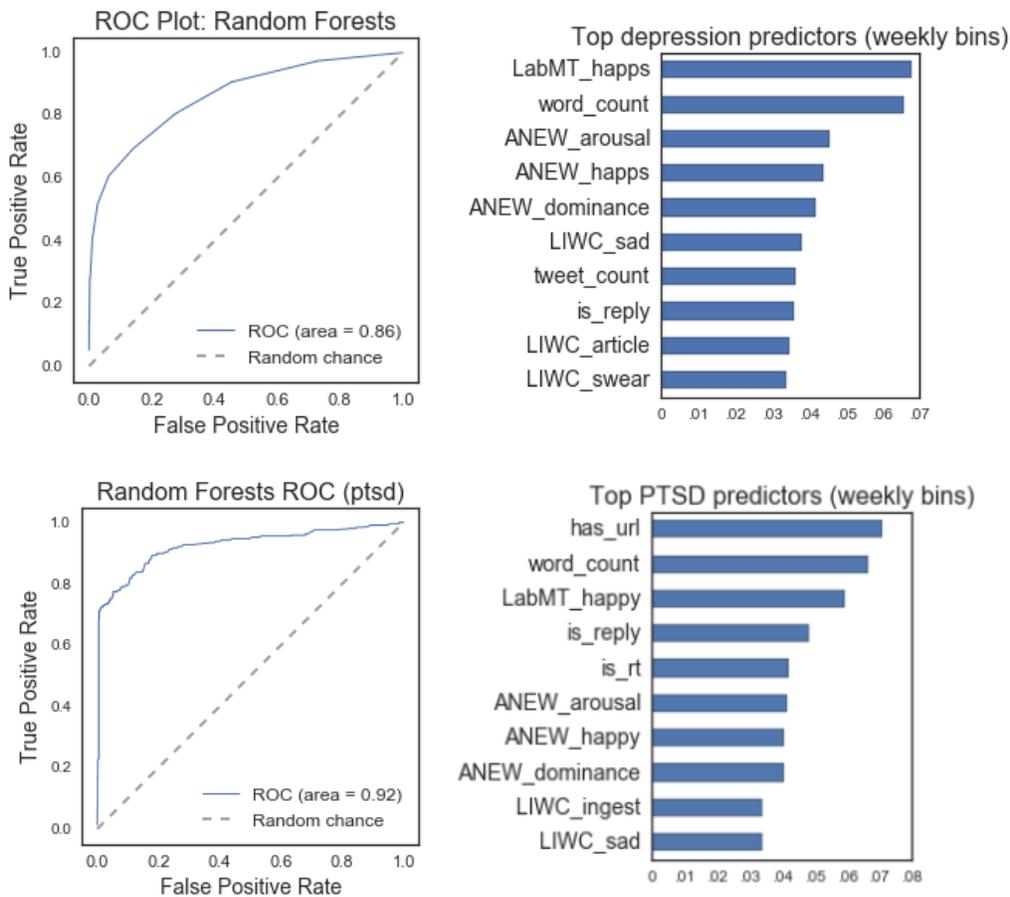

Figure SI 1. ROC curve and top predictors for Random Forests algorithm using weekly units of observation, for depression and PTSD samples ($N_{depr}$=29,328, $N_{ptsd}$=12,676). Predictor names ending in "_happy" are happiness measures; LIWC predictors (36) refer to the occurrence of semantic categories (eg. LIWC_ingest refers to food and eating words, LIWC_swear refers to profanity).

III. Random Forests hyperparameter optimization

Random Forests parameters were optimized using stratified five-fold cross-validation. The optimization schema is the same as used in Reece and Danforth (12). The optimization routine traversed every combination over the following values (best performing values are highlighted above in red):

>n_estimators = [120, 300, 500, 800, 1200]
>max_depth = [5, 8, 15, 25, 30, None]
>min_samples_split = [1, 2, 5, 10, 15, 100]
>min_samples_leaf = [1, 2, 5, 10]
>max_features = ['log2', 'sqrt', None]

IV. PTSD word shift

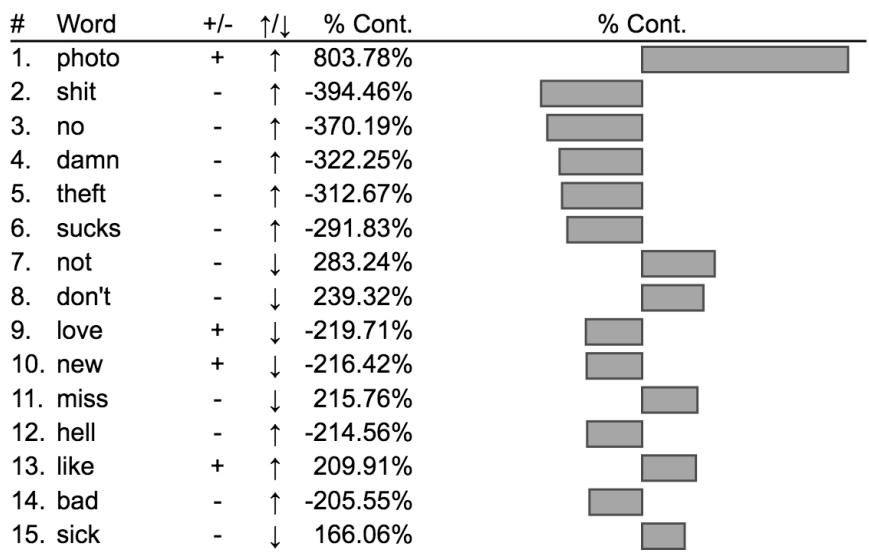

| #   | Word  | +/- | ↑/↓ | % Cont.   |
|-----|-------|-----|-----|-----------|
| 1.  | photo | +   | ↑   | 803.78%   |
| 2.  | shit  | -   | ↑   | -394.46%  |
| 3.  | no    | -   | ↑   | -370.19%  |
| 4.  | damn  | -   | ↑   | -322.25%  |
| 5.  | theft | -   | ↑   | -312.67%  |
| 6.  | sucks | -   | ↑   | -291.83%  |
| 7.  | not   | -   | ↓   | 283.24%   |
| 8.  | don't | -   | ↓   | 239.32%   |
| 9.  | love  | +   | ↓   | -219.71%  |
| 10. | new   | +   | ↓   | -216.42%  |
| 11. | miss  | -   | ↓   | 215.76%   |
| 12. | hell  | -   | ↑   | -214.56%  |
| 13. | like  | +   | ↑   | 209.91%   |
| 14. | bad   | -   | ↑   | -205.55%  |
| 15. | sick  | -   | ↓   | 166.06%   |

Figure SI 2. PTSD word-shift graph revealing contributions to Twitter happiness observed for PTSD (6.10) and healthy (6.10) participants. In column 3, (-) indicates a relatively negative word, and (+) indicates a relatively positive word, both with respect to the average happiness of all healthy tweets. An up (down) arrow indicates that word was used more (less) by the PTSD class. Words on the left (right) contribute to a decrease (increase) in happiness in the PTSD class. In column 5, % contribution is calculated with respect to the overall average happiness difference between PTSD and healthy participants, which was quite small.

23